\documentclass[12pt]{iopart}%
\usepackage{iopams}
\usepackage{graphicx}%
\begin{document}

\title[Integrability of Bianchi Universes]{Integrability of anisotropic and homogeneous Universes in scalar-tensor theory of gravitation}
\author{Julien\ Larena}
\address{Laboratoire Univers et Th\'eories, UMR8102, Paris
    Observatory and University Paris 7 Denis Diderot, julien.larena@obspm.fr} 
\author{J\'erome \ Perez}
\address{Laboratoire de Math\'ematiques Appliqu\'ees, Ecole Nationale Sup\'erieure de Techniques Avanc\'ees, jerome.perez@ensta.fr}

\begin{abstract}
In this paper, we develop a method based on the analysis of the Kovalewski
exponents to study the integrability of anisotropic and homogeneous Universes.
The formalism is developed in scalar-tensor gravity, the
general relativistic case appearing as a special case of this larger
framework.
Then, depending on the rationality of the Kovalewski exponents, the different
models, both in the vacuum and in presence of a barotropic matter fluid, are
classified, and their integrability is discussed.
\end{abstract}

Greek indexes run from 0 to 4, Latin indexes run from 1 to 3 and
$\nabla^{\mu}$ indicates a covariant derivative.
\pacs{02.30.Ik, 04.20.-q, 04.20.Cv, 98.80.Jk}
\maketitle
\section{\bigskip Dynamical equations}

\subsection{Homogeneous spaces}

The classification of homogeneous and anisotropic $3D$ spaces gives $11$ different
types of spaces associated to distinct families of structure group constants
\begin{equation}
C_{ab}^{\;c}=\varepsilon_{abd}\,N^{dc}+\delta_{b}^{c}\,A_{a}-\delta_{a}%
^{c}\,A_{b} \label{decomp}%
\end{equation}
where $\varepsilon_{abd}\,$is the usual totally antisymmetric unit tensor,
$\delta_{b}^{c}$ the Kronecker symbol, $N^{ab}$ is the contravariant component
of an order 2 symmetric tensor, and the vector $A$ must follows
\begin{equation}
N^{ab}A_{b}=N_{ab}A^{b}=0 \label{jaca} \mbox{.}%
\end{equation}
Without loss generality, one can write $N^{ab}=\mathrm{diag}(n_{1}%
,n_{2},n_{3})$ and $A_{b}=\left[  a,0,0\right]  $ provided that $an_{1}=0$.
Distinct homogeneous and anisotropic spaces in $3D$ can then be classified in
the following table:

\begin{center}%
\begin{tabular}
[c]{c|ccc|c|c|}
& $n_{1}$ & $n_{2}$ & $n_{3}$ & $a$ & Name\\\hline\hline
$0$ is eigenvalue of $N$ with multiplicity 3 & $0$ & $0$ & $0$ & 0 &
$B_{\mbox{\textsc{i}}}$\\
& $0$ & $0$ & $0$ & $\forall$ & $B_{\mbox{\textsc{v}}}$\\\hline\hline
$0$ is eigenvalue of $N$ with multiplicity 2 & $1$ & $0$ & $0$ & 0 &
$B_{\mbox{\textsc{ii}}}$\\
& $0$ & $1$ & $0$ & $\forall$ & $B_{\mbox{\textsc{iv}}}$\\\hline\hline
$0$ is eigenvalue of $N$ with multiplicity 1 & $1$ & $1$ & $0$ & 0 &
$B_{\mbox{\textsc{vii}}_{o}}$\\
& $0$ & $1$ & $1$ & $\forall$ & $B_{\mbox{\textsc{vii}}_{a}}$\\
& $1$ & $-1$ & $0$ & $0$ & $B_{\mbox{\textsc{vi}}_{o}}$\\
& $0$ & $1$ & $-1$ & $\neq1$ & $B_{\mbox{\textsc{vi}}_{a}}$\\
& $0$ & $1$ & $-1$ & $1$ & $B_{\mbox{\textsc{iii}}}$\\\hline\hline
$0$ is not an eigenvalue of $N$ & $1$ & $1$ & $1$ & $0$ &
$B_{\mbox{\textsc{ix}}}$\\
& $1$ & $1$ & $-1$ & $0$ & $B_{\mbox{\textsc{viii}}}$%
\end{tabular}
\label{tableau}
\end{center}

Note that each case is degenerate, for example $n_{1}=-1,n_{2}=1$, $n_{3}=-1 $
and $a=0$ is in the equivalence class of B$_{\mbox{\textsc{ix}}}$ which
contains all the possibilities with a positive signature and $0$ out of the spectrum
of $N$. This is the well known Bianchi classification (see \cite{Bianchi
1},\cite{Bianchi2}).

\subsection{Scalar-tensor theory of gravitation}

In scalar-tensor theories of gravity, the dynamics of the Universe contains a
new scalar degree of freedom that couples explicitly to the energy content of
the Universe \cite{Brans,Bergmann,Nord,Wagoner,Espo}. In units of $c=1$, the
action generically writes, in the so-called Einstein frame:
\begin{eqnarray}
S  &  =\frac{1}{4\pi G}\int\left(  \frac{R}{4}- \frac{1}{2}\varphi_{,\mu
}\varphi^{,\mu} - U(\varphi)\right)  \sqrt{-g}d^{4}x\nonumber\\
&  + S_{m}(\psi_{m},\Theta^{2}(\varphi)g_{\mu\nu}), \label{eq:EFaction}%
\end{eqnarray}
$G$ being a bare gravitational constant, $\varphi$ the scalar field,
$U(\varphi)$ its self-interaction term and $\Theta(\varphi)$ its coupling to
matter. The functional $S_{m}(\psi_{m},\Theta^{2}(\varphi)g_{\mu\nu})$ stands
for the action of any field $\psi_{m}$ that contributes to the energy content
of the Universe. It expresses the fact that all these fields couple
universally to a conformal metric $\tilde{g}_{\mu\nu}=\Theta^{2}%
(\varphi)g_{\mu\nu}$, then implying that the weak equivalence principle (local
universality of free fall for non-gravitationally bound objects) holds in this
class of theories. The metric $\tilde{g}_{\mu\nu}$ defines the Dicke-Jordan
frame, in which standard rods and clocks can be used to make measurements
(since in this frame, the matter part of the action acquires its standard
form). Despite the conformal relation, these two frames have a different
status: in the Dicke-Jordan frame, where the gravitational degrees of freedom
are mixed, the Lagrangian for the matter fields does not contain explicitly
the new scalar field: the non gravitational physics has then its standard
form. In the Einstein frame, the scalar degree of freedom explicitly couples
to the matter fields, then leading for example to the variation of the
inertial masses of point-like particles. Of course, the two frames describe
the same physical world. Nevertheless, the usual interpretation of the
observable quantities is profoundly modified in the Einstein frame, whereas it
holds in the Dicke-Jordan frame, where the rods and clocks made with matter
are not affected by the presence of the scalar field. That is why one usually
refers to the Dicke-Jordan frame as the observable one. However, the
dynamics of the fields is generally more easily described in the Einstein
frame, so that in this work, since we are interested in the integrability of
the models rather than in there physical content, the analysis will be done in
the Einstein frame.

Varying the Einstein frame action (\ref{eq:EFaction}) with respect to the
fields yields the equations:
\begin{eqnarray}
\label{fieldseq}
R_{\mu\nu}-\frac{1}{2}Rg_{\mu\nu} =\chi T_{\mu\nu}+T_{\mu\nu}^{\varphi
}& &\\
\Box\varphi=-\frac{\chi}{2}\omega(\varphi)T+\frac{dU(\varphi)}{d\varphi}& &\\
\nabla_{\nu}T_{\mu}^{\nu}=\omega(\varphi)T\nabla_{\mu}\varphi& &
\end{eqnarray}
where $\chi=8\pi G,$ $T$ is the trace of the energy-momentum tensor of matter
fields $T_{\mu\nu}$, and $T_{\mu\nu}^{\varphi}=2\varphi_{,\mu}\varphi_{,\nu
}-g_{\mu\nu}(g^{\alpha\beta}\varphi_{\alpha}\varphi_{\beta})-2U(\varphi
)g_{\mu\nu}$ is the energy-momentum tensor of the scalar field.
Moreover, we have defined  the coupling $\omega\left(  \varphi\right)=\frac{d\ln\Theta
}{d\varphi}$. It is important to note that these equations reduce to those of
General Relativity in presence of a scalar field iff $\omega(\varphi)=0$.

\subsection{Dynamical equations of scalar-tensor theory in homogeneous spaces}

If one denotes by $t$ the physical time and by $\tau$ a conformal time such
that $dt=N\left(  \tau\right)  d\tau$, where $N\left(  \tau\right)  $ is
generally called \textquotedblright lapse function\textquotedblright, the line
element of the $3+1$ physical space reads:
\begin{equation}
ds^{2}=g_{ij}dx^{i}dx^{j}-dt^{2}=\gamma\,\left(  \tau\right)
\omega^{i}\omega^{j}-N^{2}\left(  \tau\right) d\tau^{2}%
\end{equation}
As explained in \cite{macallum}, one can find an invariant basis of differential
forms $\left\{  \omega^{1},\omega^{2},\omega^{3}\right\}  $ in each Bianchi
space such that $\gamma$ is diagonal; we note hereafter $\gamma\,\left(
.\right)  =\mathrm{diag}\left(  \alpha_{i}\left(  .\right)  \right)  .$

Taking $V\left(  \tau\right)  =\left(  \alpha_{1}\alpha_{2}\alpha_{3}\right)
^{1/2}$ as a lapse function, and introducing $A_{i}=\ln\left(  \alpha
_{i}\right)  $, the scalar-tensor dynamical equations in homogeneous and
anisotropic Universe read:
\begin{equation}
\left\{
\begin{array}
[c]{rll}%
\chi V^{2}\left[  3\left(  P+P_{\varphi}\right)  +\rho+\rho_{\varphi}\right]
& = & -\left(  \ln\left(  V^{2}\right)  \right)  ^{\prime\prime}+\frac{1}%
{2}\left(  \left(  \ln\left(  V^{2}\right)  \right)  ^{\prime}\right)
^{2}\nonumber\\
 & & -\frac{1}{2}\left(  A_{1}^{\prime2}+A_{2}^{\prime2}+A_{3}^{\prime2}\right)\\
\chi V^{2}\left[  \rho+\rho_{\varphi}-\left(  P+P_{\varphi}\right)  \right]  &
= & A_{1}^{\prime\prime}+n_{1}^{2}\alpha_{1}^{2}-\left(  n_{2}\alpha_{2}%
-n_{3}\alpha_{3}\right)  ^{2}\\
\chi V^{2}\left[  \rho+\rho_{\varphi}-\left(  P+P_{\varphi}\right)  \right]  &
= & A_{2}^{\prime\prime}+n_{2}^{2}\alpha_{2}^{2}-\left(  n_{3}\alpha_{3}%
-n_{1}\alpha_{1}\right)  ^{2}\\
\chi V^{2}\left[  \rho+\rho_{\varphi}-\left(  P+P_{\varphi}\right)  \right]  &
= & A_{3}^{\prime\prime}+n_{3}^{2}\alpha_{3}^{2}-\left(  n_{1}\alpha_{1}%
-n_{2}\alpha_{2}\right)  ^{2}\\
\chi V^{2}\omega\left(  \varphi\right)  \left[  \rho-3P\right]  & = &
-2\varphi^{\prime\prime}-4\left(  \ln V\right)  ^{\prime}\varphi^{\prime
}-2\frac{dU}{d\varphi}V^{2}%
\end{array}
\right.  \label{dynamiquesanshyp}%
\end{equation}

In these equations we have written $^{\prime}$ for $d/d\tau$; in addition the
Universe is filled by a perfect fluid with pressure $P$ and energy density
$\rho$; finally we have noted:
\begin{equation}
\left\{
\begin{array}
[c]{rll}%
\rho_{\varphi} & := & \left[  \varphi^{\prime2}/2+U(\varphi)\right]  /\chi\\
P_{\varphi} & := & \left[  \varphi^{\prime2}/2-U(\varphi)\right]  /\chi
\end{array}
\right.\label{vdelta2}
\end{equation}

Taking $U\equiv0$, and reorganizing the first dynamical equation using the
three others, the system (\ref{dynamiquesanshyp}) becomes:
\begin{equation}
\left\{
\begin{array}
[c]{rll}%
0 & = & E_{c}+E_{p}-4\chi\rho V^{2}-4\chi\rho_{\varphi
}V^{2}\\
\chi V^{2}\left[  \rho+\rho_{\varphi}-\left(  P+P_{\varphi}\right)  \right]  &
= & A_{1}^{\prime\prime}+n_{1}^{2}\alpha_{1}^{2}-\left(  n_{2}\alpha_{2}%
-n_{3}\alpha_{3}\right)  ^{2}\\
\chi V^{2}\left[  \rho+\rho_{\varphi}-\left(  P+P_{\varphi}\right)  \right]  &
= & A_{2}^{\prime\prime}+n_{2}^{2}\alpha_{2}^{2}-\left(  n_{3}\alpha_{3}%
-n_{1}\alpha_{1}\right)  ^{2}\\
\chi V^{2}\left[  \rho+\rho_{\varphi}-\left(  P+P_{\varphi}\right)  \right]  &
= & A_{3}^{\prime\prime}+n_{3}^{2}\alpha_{3}^{2}-\left(  n_{1}\alpha_{1}%
-n_{2}\alpha_{2}\right)  ^{2}\\
\chi V^{2}\omega\left(  \varphi\right)  \left[  \rho-3P\right]  & = &
-2\varphi^{\prime\prime}-4\left(  \ln V\right)  ^{\prime}\varphi^{\prime}
\end{array}
\right. \label{systeme}
\end{equation}
where
\[
E_{c}:=A_{1}^{\prime}A_{2}^{\prime}+A_{1}^{\prime}A_{3}^{\prime}+A_{3}%
^{\prime}A_{2}^{\prime} \;\;\;\mbox{and}\;\;\;
E_{p}:=\sum\limits_{i\neq j=1}^{3}n_{i}n_{j}e^{A_{i}+A_{j}}-\sum
\limits_{i=1}^{3}n_{i}^{2}e^{2A_{i}} 
\]
The energy-momentum conservation, that in scalar-tensor theory is:
\begin{equation}
\nabla^{\mu}T_{\mu\nu}=\omega\left(  \varphi\right)  T_{\lambda\beta
}g^{\lambda\beta}\partial_{\nu}\varphi
\end{equation}
coupled to the hypothesis of a barotropic fluid:
\begin{equation}
P=\left(  \Gamma-1\right)  \rho
\end{equation}
allows us to obtain a relation between $\rho,$ $V$ and $\varphi$ which is:
\begin{equation}
\rho=\rho_{o}V^{-\Gamma}\Theta^{4-3\Gamma}\ \ \ \ \ \ \ \ \ \mbox{ with }%
\rho_{o}\in \mathbb{R}^{+}\setminus\left\{  0\right\}  \label{rhovphi}%
\end{equation}
One should note that a fluid of radiation $(\Gamma=4/3)$ doesn't couple directly to the scalar field. Assuming a power law dependence of
$\varphi^{\prime}$ in $V$, more precisely:
\begin{equation}
\label{assump}
\varphi^{\prime}=\varphi^{\prime}_{o}V^{\Delta} \label{vdelta}
\end{equation}
and making use of (\ref{rhovphi}) one can solve the last equation of system
(\ref{dynamiquesanshyp}), and obtain an explicit dependence on $V$ for
$\Theta(\varphi)$:
\begin{equation}
\label{thetaV}\rho_{o} \Theta^{4-3\Gamma}= \frac{a^2}{2\chi} V^{2(\Delta-1)+\Gamma}\:\:\:\mbox{where}\;a^2=
\frac{4(\Delta+2)}{2(1-\Delta)-\Gamma}\varphi^{\prime 2}_{o}
\end{equation}
Assuming $\rho_{o} >0$ (which corresponds to non exotic matter) and
$0\le\Gamma\le2$ (which corresponds to the largest class of barotropic fluid),
relation (\ref{thetaV}) holds under the condition: 
\begin{equation}
-2 < \Delta <\frac{2-\Gamma}{2}
\end{equation}

The assumption (\ref{assump}) is a constraint on the entire dynamical system
rather than on the scalar-tensor theory itself. Indeed, any choice of coupling
function $\omega(\varphi)$ can be done, but then the resulting behavior of
$V(\tau)$ is completely fixed by the last equation of system
(\ref{systeme}). Conversely, imposing a behavior for the volume $V(\tau)$
determines the corresponding coupling function. For example, imposing a
Brans-Dicke theory, i.e. $\omega(\varphi)=\omega_{0}=cste$ results in
$V(\tau)\propto \tau^{1/(2-\Delta)}$; on the contrary, a volume evolving as
$V\propto e^{\lambda\tau}$ leads to   
\begin{equation}
\Theta(\varphi)\propto\varphi^{\frac{2(\Delta-1)+\Gamma}{(4-3\Gamma)\Delta}}\mbox{
,}
\end{equation}
that is,  $\omega(\varphi)\propto\frac{1}{\varphi}$ and $\varphi\propto
e^{\lambda\Delta\tau}$.
To sum up, the only constraint imposed by assumption (\ref{assump}) is on the
couple $(V(\tau),\omega(\varphi))$, through the relation:
\begin{equation}
\frac{\rho}{\rho_{\varphi}}=\frac{4(\Delta+2)}{2(1-\Delta)-\Gamma}V^{-2}\mbox{
,}
\end{equation}
that implies that, at any time, the ratio of the densities of the barotropic
fluid and of the scalar field scales with the inverse of the square of the volume.
This is a sufficient constraint to make the Kovalewski formalism tractable in
scalar-tensor gravity.
Replacing (\ref{thetaV}) into (\ref{rhovphi}), and considering  (\ref{vdelta}) after (\ref{vdelta2}), the dynamical system associated with the homogeneous
Universe in scalar-tensor theory is:
\begin{equation}
\left\{
\begin{array}
[c]{rll}%
E_{c}+E_{p} & = & 2a^{2} V^{2\Delta}+2\varphi^{\prime\;2}_o V^{2\Delta+2}\\
2\left[  2-\Gamma\right]  a^2 V^{2\Delta} & = & A_{1}^{\prime\prime
}+n_{1}^{2}\alpha_{1}^{2}-\left(  n_{2}\alpha_{2}-n_{3}\alpha_{3}\right)
^{2}\\
2\left[  2-\Gamma\right]  a^2 V^{2\Delta} & = & A_{2}^{\prime\prime
}+n_{2}^{2}\alpha_{2}^{2}-\left(  n_{3}\alpha_{3}-n_{1}\alpha_{1}\right)
^{2}\\
2\left[  2-\Gamma\right]  a^2 V^{2\Delta} & = & A_{3}^{\prime\prime
}+n_{3}^{2}\alpha_{3}^{2}-\left(  n_{1}\alpha_{1}-n_{2}\alpha_{2}\right)  ^{2}%
\end{array}
\right.  \label{dynsyst}%
\end{equation}
It is important to note that whereas this system seems independent on $\Gamma
$, it actually depends on it through the parameter $\Delta$ that fully
characterizes our solution. In fact, for each coupling function, their exists a non-ambiguous link between
$\Gamma$ and $\Delta$. For example in the case of a radiation fluid when $\Gamma=4/3$, the LHS of the last equation in system (\ref{systeme}) vanishes and one can find that $\varphi'\propto V^{-2}$ and then $\Delta=-2$. 

A direct inspection of the first equation of system (\ref{dynsyst}) shows that
we can predict qualitatively the behavior of the case: $\Delta=0$. Indeed, in this case, one of the two terms of
$E_{c}+E_{p}$ reduces to a positive constant and the other one tends to $0$ as
$V$ tends to $0$. It is well known (e.g. \cite{BK}) that such a case breaks
the Kasner cycle (e.g. \cite{BKL}) for $B_{\mbox{\textsc{viii}}}$ and $B_{\mbox{\textsc{ix}}}$, and then suppresses
the chaotic behaviour toward the $t=0$ singularity for these models. The case
$\Delta=-1$ seems to be similar, but the other term now diverges and then this
simple analysis cannot be done. 

\subsection{Hamiltonian formalism}

The quantity called $\ E_{c}=A_{1}^{\prime}A_{2}^{\prime}+A_{1}^{\prime}%
A_{3}^{\prime}+A_{3}^{\prime}A_{2}^{\prime}$ is a quadratic form of
$A_{i=1,2,3}$ derivatives. Then, one can diagonalize it using a linear change
of variables:
\begin{equation}
\left\{
\begin{array}
[c]{l}%
q_{1}= \left(  A_{1}-A_{2} \right)  /\sqrt{2}\\
q_{2}= \left(  A_{1}+A_{2}-2A_{3}\right)  /\sqrt{6}\\
q_{3}=2\left(  A_{1}+A_{2}+ A_{3}\right)  /\sqrt{6}%
\end{array}
\right.  \label{changvar}%
\end{equation}
Introducing the associated conformal time derivatives $p_{i=1,2,3}%
:=q_{_{i=1,2,3}}^{\prime}$, the first equation of the system (\ref{dynsyst}) becomes:
\begin{equation}
H:=\frac{1}{2}\left(  p_{3}^{2}-p_{1}^{2}-p_{2}^{2}\right)  +e^{\frac{\sqrt{6}%
}{3}q_{3}}\,\xi\left(  q_{1},q_{2}\right)  -2a^2 e^{\frac{\sqrt{6}}{2}\Delta
q_{3}}-2\varphi_{o}^{\prime2}e^{\frac{\sqrt{6}}{2}\left(  \Delta+1\right)  q_{3}}=0
\label{hamiltonian}%
\end{equation}
the so-called potential $\xi$ is defined by:
\begin{eqnarray}
\xi\left(  q_{1},q_{2}\right)   &  =-n_{1}^{2}e^{\left(  \frac{\sqrt{6}}%
{3}q_{2}+\sqrt{2}q_{1}\right)  }-n_{2}^{2}e^{\left(  \frac{\sqrt{6}}{3}%
q_{2}-\sqrt{2}q_{1}\right)  }-n_{3}^{2}e^{-\frac{2\sqrt{6}}{3}q_{2}%
}\label{xiqq}\\
&  +2n_{1}n_{2}e^{\frac{\sqrt{6}}{3}q_{_{2}}}+2n_{1}n_{3}e^{\left(
\frac{\sqrt{2}}{2}q_{1}-\frac{\sqrt{6}}{6}q_{2}\right)  }+2n_{2}%
n_{3}e^{-\left(  \frac{\sqrt{2}}{2}q_{1}+\frac{\sqrt{6}}{6}q_{2}\right)
}\nonumber
\end{eqnarray}

In terms of $\left(  q_{i},p_{i}\right)  $ variables, the dynamical system
$\left(  \ref{dynsyst}\right)  $ is quasi-Hamiltonian (e.g. \cite{misner1} and later \cite{Misner2}%
,\cite{ryan},\cite{Jantzen})
\begin{equation}%
\begin{array}
[c]{cc}%
q_{1,2}^{\prime}=\frac{dq_{1,2}}{d\tau}=-\frac{\partial H}{\partial p_{1,2}}
& p_{1,2}^{\prime}=\frac{dp_{1,2}}{d\tau}=-\frac{\partial H}{\partial
q_{1,2}}\\
& \\
q_{3}^{\prime}=\frac{dq_{3}}{d\tau}=\frac{\partial H}{\partial p_{3}} &
p_{3}^{\prime}=\frac{dp_{3}}{d\tau}=-\frac{\partial H}{\partial q_{3}}%
\end{array}
\label{eqdyn}%
\end{equation}
The minus that breaks the strict Hamiltonian symmetry comes from the minus
signature of the quadratic form associated with $E_{c}$.

\section{Integrability of homogeneous Universes}

\subsection{The case of Bianchi Universes}

The following work has been initiated by Melnikov's team (see e.g. \cite{Melnikov}
 and \cite{kirilovmelnikov} and references therein). In the special case of $B_{\textsc{ix}}$,
\cite{pavlov} consists in an application; a generalization in the context of
 the whole class $A$ (i.e. $a=0$) was tried by \cite{polonais}. However, a lot of
imprecisions in this last work need this new reformulation and extension to
scalar-tensor theory.

\subsubsection{Bianchi universes as generalized Toda systems}

Introducing the following vectors:
\begin{equation}%
\fl
\begin{array}
[c]{lll}%
\mathbf{a}_{1} :=[                   0,  \sqrt{6}/{3},  \sqrt{6}/{3} ]    & 
\mathbf{a}_{2} :=[  \sqrt{2}/{2}      ,- \sqrt{6}/{6},  \sqrt{6}/{3} ]    & 
\mathbf{a}_{3} :=[- \sqrt{2}/{2},      - \sqrt{6}/{6},  \sqrt{6}/{3} ]      \\
\mathbf{a}_{4} :=[       \sqrt{2}    ,   \sqrt{6}/{3},  \sqrt{6}/{3} ]    &
\mathbf{a}_{5} :=[      -\sqrt{2}    ,   \sqrt{6}/{3},  \sqrt{6}/{3} ]    &
\mathbf{a}_{6} :=[                  0,- 2\sqrt{6}/{3},  \sqrt{6}/{3} ]       \\
\mathbf{a}_{7} :=[                  0,                   0,
  \sqrt{6}\Delta/{2}] & &                                                                                   
\mathbf{a}_{8} :=[                  0,                   0,  \sqrt{6}
  (\Delta+1)/{2}]
\end{array}
\label{liste}%
\end{equation}
the 3-forms:
\begin{equation}
\forall\mathbf{x},\mathbf{y}\in\mathbb{R}^{3}\;\;
\begin{array}
[c]{c}%
\left(  x,y\right)  :=+x_{1}y_{1}+x_{2}y_{2}+x_{3}y_{3}\\
\left\langle x,y\right\rangle :=-x_{1}y_{1}-x_{2}y_{2}+x_{3}y_{3}%
\end{array}
\end{equation}
and the constants:
\begin{equation}%
\begin{array}
[c]{ccc}%
k_{1}:=2n_{1}n_{2} & k_{2}:=2n_{1}n_{3} & k_{3}:=2n_{2}n_{3}\\
k_{4}:=-n_{1}^{2} & k_{5}:=-n_{2}^{2} & k_{6}:=-n_{3}^{2}\\
k_{7}=-2a^{2} & k_{8}=-2\varphi_{o}^{\prime2} & %
\end{array}
\end{equation}
it is clear that the Hamiltonian $\left(  \ref{hamiltonian}\right)  $ reads:
\begin{equation}
H=\frac{1}{2}\left\langle \mathbf{p},\mathbf{p}\right\rangle +\sum_{i=1}%
^{8}k_{i}e^{\left(  \mathbf{a}_{i},\mathbf{q}\right)  }%
\end{equation}
which is a classical form of a generalized Toda dynamical system.

Following Melnikov \cite{Melnikov}, we change $\left\{  \mathbf{q}%
,\mathbf{p}\right\}  $ variables to $\left\{  \mathbf{u},\mathbf{v}\right\}  $
ones, such that:
\begin{equation}
\left\{
\begin{array}
[c]{c}%
\mathbf{u}\in\mathbb{R}^{8}\mbox{, }u_{i=1,...,8}:=\left\langle \mathbf{a}%
_{i},\mathbf{p}\right\rangle \\
\mathbf{v}\in\mathbb{R}^{8}\mbox{, }v_{i=1,...,8}:=\exp\left(  \mathbf{a}%
_{i},\mathbf{q}\right)
\end{array}
\right.
\end{equation}
Through this change, the number of degrees of freedom jumps from 6 in $\left\{
\mathbf{q},\mathbf{p}\right\}  $ to 16 in terms of $\left\{  \mathbf{u}%
,\mathbf{v}\right\}  $. Still writing $^{\prime}$ for the derivatives with
respect to the conformal time $\tau$, the dynamical equations then read:
\begin{equation}
\forall i=1,...,8\,\;\;\;\;\left\{
\begin{array}
[c]{l}%
v_{i}^{\prime}=u_{i}v_{i}\\
\\
u_{i}^{\prime}=%
{\displaystyle\sum\limits_{j=1}^{8}}
m_{ij}v_{j}%
\end{array}
\right.  \;\;\;\;\;\;\;\mbox{with }m_{ij}:=-k_{j}\left\langle \mathbf{a}%
_{i},\mathbf{a}_{j}\right\rangle \label{syspoly}%
\end{equation}
This new formulation is now polynomial. Using the appendix notations, one can
directly prove that the system $\left(\ref{syspoly}\right)$ is autosimilar with any non vanishing
index and weight $\mathbf{g}$ such that:
\begin{equation}
g_{1}=...=g_{8}=1\mbox{ \thinspace\thinspace\thinspace\thinspace
\thinspace\thinspace and \thinspace\thinspace\thinspace\thinspace
\thinspace\thinspace}g_{9}=...=g_{16}=2
\end{equation}
$\mathbf{g}$ is unique provided that $\left(  \ref{condunicite}\right)  $ is fulfilled.
A particular autosimilar solution of $\left(  \ref{syspoly}\right)  $ is then
\begin{equation}
\mathbf{[\tilde{u}\mbox{ } \tilde{v}]^{T}}=\mathbf{c}t^{-\mathbf{g}}=\left[  \lambda_{1}%
t^{-1},...,\lambda_{8}t^{-1},\mu_{1}t^{-2},...,\mu_{8}t^{-2}\right]  ^{T}%
\end{equation}
provided that the constant non vanishing vector $\mathbf{c}$ $=\left[
\mathbf{\lambda},\mathbf{\mu}\right]  =\left[  \lambda_{1},...,\lambda_{8}%
,\mu_{1},...,\mu_{8}\right]  $ is a solution of the algebraic system of
equations:
\begin{equation}
\forall i=1,\cdots,8\;\; \left\{
\begin{array}
[c]{l}%
{\displaystyle\sum\limits_{j=1}^{8}} m_{ij}\,\mu_{j}=-\lambda_{i}\\
\\
\lambda_{i}\,\mu_{i}=-2\mu_{i}%
\end{array}
\right.  \label{sysalgeb}%
\end{equation}
Hence, to any non vanishing solution of this last system corresponds a set of
16 Kovalewski exponents that allows to write the solution of the system
(\ref{eqdyn}) (cf Appendix A). 
A necessary condition for the system to be integrable is that all its
Kovalewski exponents be rational. 
So, the rest of the paper will be devoted to the analysis of these exponents in order to study the
integrability of different types of homogeneous and anisotropic Universes.

\subsubsection{Solutions of the algebraic system}

In the more general case (that is scalar-tensor theory in presence of matter
barotropic fluids), the algebraic system $\left(  \ref{sysalgeb}\right)  $
makes use of the $8\times8$ matrix
\begin{equation}
\fl
M :=\left[  \mbox{ {\small
\begin{tabular}
[c]{cccccccc}%
0 & $-2n_{1} n_{3}$ & $-2n_{3} n_{2}$ & 0 & 0 & $2n_{3}^{2}$ & $2a^{2} \Delta$
& $2\varphi^{\prime\;2}_o  \Delta_1$\\
$-2n_{1} n_{2}$ & 0 & $-2n_{3} n_{2}$ & 0 & $2n_{2}^{2}$ & 0 & $2a^{2} \Delta$
& $2\varphi^{\prime\;2}_o  \Delta_1$\\
$-2n_{1} n_{2}$ & $-2n_{1} n_{3}$ & 0 & $2n_{1}^{2}$ & 0 & 0 & $2a^{2} \Delta$
& $2\varphi^{\prime\;2}_o  \Delta_1 $\\
0 & 0 & $-4n_{3} n_{2}$ & $-2n_{1}^{2}$ & $2n_{2}^{2}$ & $2n_{3}^{2}$ & $2a^{2}
\Delta$ & $2\varphi^{\prime\;2}_o  \Delta_1$\\
0 & $-4n_{1} n_{3}$ & 0 & $2n_{1}^{2}$ & $-2n_{2}^{2}$ & $2n_{3}^{2}$ & $2a^{2}
\Delta$ & $2\varphi^{\prime\;2}_o \Delta_1$\\
$-4n_{1} n_{2}$ & $-4n_{1} n_{3}$ & 0 & $2n_{1}^{2}$ & $2n_{2}^{2}$ &
$-2n_{3}^{2}$ & $2a^{2} \Delta$ & $2\varphi^{\prime\;2}_o  \Delta_1$\\
$-2n_{1} n_{2} \Delta$ & $-2n_{1} n_{3} \Delta$ & $-2n_{3} n_{2} \Delta$ &
$n_{1}^{2} \Delta$ & $n_{2}^{2} \Delta$ & $n_{3}^{2} \Delta$ & $ 3  a^{2}
\Delta^{2} $ & $3 \varphi^{\prime\;2}_o  \Delta_1 \Delta$\\
$-2n_{1} n_{2} \Delta_1 $ & $-2n_{1} n_{3} \Delta_1$ & $-2n_{3} n_{2}
\Delta_1 $ & $n_{1}^{2} \Delta_1 $ & $n_{2}^{2} \Delta_1 $ & $n_{3}^{2}
\Delta_1 $ & $ 3  a^{2} \Delta\Delta_1 $ & $3 \varphi^{\prime\;2}_o 
\Delta_1^{2}$\\
&  &  &  &  &  &  &
\end{tabular}
}} \right]  \label{matriceM}%
\end{equation}
where $\Delta_1 :=\Delta+1$. One can straightforwardly check that 
$M$ is  of rank 3.

The vector $\stackrel{\rightarrow}{0}$ is always a solution of the system (\ref{sysalgeb}), but it is not
relevant for the quest of Kovalewski exponents that needs non trivial solutions.
A systematic study of the solutions of the system (\ref{sysalgeb}) is possible : for
$p=6,7$ or $8$ let $E_{p}=\left\{ 1,2,3,\cdots,p\right\} $, we consider the
minor determinants:
\begin{equation}%
\fl
\begin{array}
[c]{cc}%
\forall(i,j,k)\in E_{p}\times E_{p}\times E_{p} & \mbox{\ \ \ \ \ }\zeta
_{i}=m_{ii}\;\;,\zeta_{i,j}=\left\vert
\begin{array}
[c]{ll}%
m_{ii} & m_{ij}\\
m_{ji} & m_{jj}%
\end{array}
\right\vert \\
& \zeta_{i,j,k}=\left\vert
\begin{array}
[c]{lll}%
m_{ii} & m_{ij} & m_{ik}\\
m_{ji} & m_{jj} & m_{jk}\\
m_{ki} & m_{kj} & m_{kk}%
\end{array}
\right\vert
\end{array}
\end{equation}
and the determinants:
\begin{eqnarray}
d_{2i}=\left\vert
\begin{array}
[c]{ll}%
2 & m_{ij}\\
2 & m_{jj}%
\end{array}
\right\vert ,\;\;d_{2j}=\left\vert
\begin{array}
[c]{ll}%
m_{ii} & 2\\
m_{ji} & 2
\end{array}
\right\vert
\\
\fl
d_{3i}=\left\vert
\begin{array}
[c]{lll}%
2 & m_{ij} & m_{ik}\\
2 & m_{jj} & m_{jk}\\
2 & m_{kj} & m_{kk}%
\end{array}
\right\vert ,\;\;d_{3j}=\left\vert
\begin{array}
[c]{lll}%
m_{ii} & 2 & m_{ik}\\
m_{ji} & 2 & m_{jk}\\
m_{ki} & 2 & m_{kk}%
\end{array}
\right\vert ,\;\;d_{3k}=\left\vert
\begin{array}
[c]{lll}%
m_{ii} & m_{ij} & 2\\
m_{ji} & m_{jj} & 2\\
m_{ki} & m_{kj} & 2
\end{array}
\right\vert
\end{eqnarray}
Solutions of System (\ref{sysalgeb}) are then classified into three classes:
\begin{enumerate}
\item Type 1 solutions (T1): $\exists!$ $i\in E_{p}$ such that $\mu_{i}\neq0$
and $\forall j\in E_{p}\setminus\left\{  i\right\}  ,$\ $\mu_{j}=0$ : then
$\lambda_{i}=-2$ and

\begin{itemize}
\item if $\zeta_{i}=0$ : there is no solution.

\item if $\zeta_{i}\neq0$ : $\mu_{i}=2/\zeta_{i}$ and $\forall j\in
E_{p}\setminus\left\{  i\right\}  ,$ $\lambda_{j}=-2m_{ji}/\zeta_{i}$
\end{itemize}

\item Type 2 solutions (T2): $\exists!$ $\left(  i,j\right)  \in E_{p}$
$\times\left(  E_{p}\setminus\left\{  i\right\}  \right)  $ such that
$\left\{  \mu_{i},\mu_{j}\right\}  \neq$ $\left\{  0,0\right\}  $ and $\forall
k\in E_{p}\setminus\left\{  i,j\right\}  ,$\ $\mu_{k}=0$ : then $\lambda
_{i}=\lambda_{j}=-2$ and

\begin{itemize}
\item if $\zeta_{i,j}=0$ : there is no solution.

\item if $\zeta_{i,j}\neq0$ : $\mu_{i}=d_{2i}/\zeta_{ij}$ and $\mu_{j}%
=d_{2j}/\zeta_{i,j}$ moreover $\forall k\in E_{p}\setminus\left\{
i,j\right\}  ,$\ $\lambda_{k}=\left(  m_{ki}d_{2i}+m_{kj}d_{2j}\right)
/\zeta_{i,j}$
\end{itemize}

\item Type 3 solutions (T3): $\exists!$ $\left(  i,j,k\right)  \in E_{p}$
$\times\left(  E_{p}\setminus\left\{  i\right\}  \right)  \times\left(
E_{p}\setminus\left\{  i,j\right\}  \right)  $ such that $\left\{  \mu_{i}%
,\mu_{j},\mu_{k}\right\}  \neq$ $\left\{  0,0,0\right\} $ and $\forall l\in
E_{p}\setminus\left\{  i,j,k\right\}  , $\ $\mu_{l}=0$ : then $\lambda
_{i}=\lambda_{j}=\lambda_{k}=-2$ and

\begin{itemize}
\item if $\zeta_{i,j,k}=0$ : there is no solution

\item if $\zeta_{i,j,k}\neq0$ : $\mu_{i}=d_{3i}/\zeta_{i,j,k}$ , $\mu
_{j}=d_{3j}/\zeta_{i,j,k}$ and $\mu_{k}=d_{3k}/\zeta_{i,j,k}$ moreover $\forall
l\in E_{p}\setminus\left\{  i,j,k\right\}  ,$\ $\lambda_{l}=\left(
m_{li}d_{3i}+m_{lj}d_{3j}+m_{lk}d_{3k}\right)  /\zeta_{i,j,k}$
\end{itemize}
\end{enumerate}

\subsubsection{Kovalewski exponents for Bianchi Universes}

In what follows, we will examine the Kovalewski exponents for Bianchi
Universes in $4$ different cases that are included in the formalism presented
above: scalar-tensor gravity with or without matter fluids, and General
Relativity with or without matter fluids. 
We proceed in three steps.
\begin{enumerate}

\item Select a Universe. It corresponds to choosing a set of $n_{i=1,2,3}$ in table
\ref{tableau}. This step determines the theory of gravitation and the matter
content of interest, and this choice is made by considering different forms
for the matrix $M$ defined in (\ref{matriceM}):

\begin{itemize}

\item Barotropic matter filled Universe in scalar-tensor theory of
  gravitation: $p=8$. This is the most general case that was presented in detail in
  the preceding subsection.

\item Empty Universe in scalar-tensor theory of gravitation: $p=7$. The associated matrix is $M$  but without the 7th line and the 7th column:
 \[
\fl
M_{STV}:=\left[  \mbox{ {\small
\begin{tabular}
[c]{ccccccc}%
0               & $-2n_{1} n_{3}$ & $-2n_{3} n_{2}$ & 0 & 0                & $2n_{3}^{2}$ & $2\varphi^{\prime\;2}_o  \Delta_1$\\
$-2n_{1} n_{2}$ & 0               & $-2n_{3} n_{2}$ & 0 & $2n_{2}^{2}$     & 0 & $2\varphi^{\prime\;2}_o  \Delta_1$\\
$-2n_{1} n_{2}$ & $-2n_{1} n_{3}$ & 0 & $2n_{1}^{2}$ & 0 & 0 
& $2\varphi^{\prime\;2}_o  \Delta_1 $\\
0 & 0 & $-4n_{3} n_{2}$ & $-2n_{1}^{2}$ & $2n_{2}^{2}$ & $2n_{3}^{2}$ & $2\varphi^{\prime\;2}_o  \Delta_1$\\
0 & $-4n_{1} n_{3}$ & 0 & $2n_{1}^{2}$ & $-2n_{2}^{2}$ & $2n_{3}^{2}$ & $2\varphi^{\prime\;2}_o \Delta_1$\\
$-4n_{1} n_{2}$ & $-4n_{1} n_{3}$ & 0 & $2n_{1}^{2}$ & $2n_{2}^{2}$ &
$-2n_{3}^{2}$ & $2\varphi^{\prime\;2}_o  \Delta_1$\\
$-2n_{1} n_{2} \Delta_1 $ & $-2n_{1} n_{3} \Delta_1$ & $-2n_{3} n_{2}
\Delta_1 $ & $n_{1}^{2} \Delta_1 $ & $n_{2}^{2} \Delta_1 $ & $n_{3}^{2}
\Delta_1 $ & $3 \varphi^{\prime\;2}_o \Delta_1^{2}$
\end{tabular}
}} \right]  \label{matriceMtilde}%
\]
\item Barotropic matter filled Universe in Einstein General Relativity: $p=7$. The associated matrix is $M$  but with the 8th line and 8th column removed, and the 7th line and 7th column adapted. Noting $\gamma$ for $2-\Gamma$ we have:
\[
\fl
M_{GRM}:=\left[  \mbox{ {\small
\begin{tabular}
[c]{ccccccc}%
0 & $-2n_{1} n_{3}$ & $-2n_{3} n_{2}$ & 0 & 0 & $2n_{3}^{2}$ & $2\rho_o \chi \gamma$\\
$-2n_{1} n_{2}$ & 0 & $-2n_{3} n_{2}$ & 0 & $2n_{2}^{2}$ & 0 & $2\rho_o \chi \gamma$\\
$-2n_{1} n_{2}$ & $-2n_{1} n_{3}$ & 0 & $2n_{1}^{2}$ & 0 & 0 & $2\rho_o \chi \gamma$\\
0 & 0 & $-4n_{3} n_{2}$ & $-2n_{1}^{2}$ & $2n_{2}^{2}$ & $2n_{3}^{2}$ & $2\rho_o \chi \gamma$\\
0 & $-4n_{1} n_{3}$ & 0 & $2n_{1}^{2}$ & $-2n_{2}^{2}$ & $2n_{3}^{2}$ & $2\rho_o \chi \gamma$\\
$-4n_{1} n_{2}$ & $-4n_{1} n_{3}$ & 0 & $2n_{1}^{2}$ & $2n_{2}^{2}$ &
$-2n_{3}^{2}$ & $2\rho_o \chi \gamma$\\
$-n_{1} n_{2} \gamma$ & $-n_{1} n_{3} \gamma$ & $-n_{3} n_{2} \gamma$ &
$n_{1}^{2} \gamma/2$ & $n_{2}^{2} \gamma/2$ & $n_{3}^{2} \gamma/2$ & $ 3\rho_o \chi \gamma^2/2$ \\
\end{tabular}
}} \right]  \label{matriceM}%
\]
\item Empty Universe in Einstein General Relativity: $p=6$. The associated matrix is $M$  but with the 7th and 8th lines and the 7th and 8th columns removed:
\[
\fl
M_{GRV}:=\left[  \mbox{ {\small
\begin{tabular}
[c]{cccccc}%
0 & $-2n_{1} n_{3}$ & $-2n_{3} n_{2}$ & 0 & 0 & $2n_{3}^{2}$\\
$-2n_{1} n_{2}$ & 0 & $-2n_{3} n_{2}$ & 0 & $2n_{2}^{2}$ & 0\\
$-2n_{1} n_{2}$ & $-2n_{1} n_{3}$ & 0 & $2n_{1}^{2}$ & 0 & 0\\
0 & 0 & $-4n_{3} n_{2}$ & $-2n_{1}^{2}$ & $2n_{2}^{2}$ & $2n_{3}^{2}$\\
0 & $-4n_{1} n_{3}$ & 0 & $2n_{1}^{2}$ & $-2n_{2}^{2}$ & $2n_{3}^{2}$\\
$-4n_{1} n_{2}$ & $-4n_{1} n_{3}$ & 0 & $2n_{1}^{2}$ & $2n_{2}^{2}$ &$-2n_{3}^{2}$
\end{tabular}
}} \right]  \label{matriceM}%
\]

\end{itemize}
We then have to apply the following algorithm to the appropriate matrix.

\item Determine all the non vanishing minor determinants $\zeta$ which could be
extracted from the considered matrix.

\item For each $\zeta$, compute the associated set of $2p$ Kovalewski
exponents. In our polynomial case, as indicated in appendix A this set is the
set of eigenvalues of the matrix:
\begin{equation}
K=\left[
\begin{array}
[c]{cccccccccc}%
1 & 0 & \cdots & \cdots & 0 & m_{11} & \cdots & \cdots & \cdots & m_{p1}\\
0 & \ddots & \ddots &  & \vdots & \vdots &  &  &  & \vdots\\
\vdots & \ddots & I_{p} & \ddots & \vdots & \vdots &  & M &  & \vdots\\
\vdots &  & \ddots & \ddots & 0 & \vdots &  &  &  & \vdots\\
0 & \cdots & \cdots & 0 & 1 & m_{1p} & \cdots & \cdots & \cdots & m_{pp}\\
\mu_{1} & 0 & \cdots & \cdots & 0 & \lambda_{1}+2 & 0 & \cdots & \cdots & 0\\
0 & \ddots & \ddots &  & \vdots & 0 & \ddots & \ddots &  & \vdots\\
\vdots & \ddots & \ddots & \ddots & \vdots & \vdots & \ddots & \ddots & \ddots
& \vdots\\
\vdots &  & \ddots & \ddots & 0 & \vdots &  & \ddots & \ddots & 0\\
0 & \cdots & \cdots & 0 & \mu_{p} & 0 & \cdots & \cdots & 0 & \lambda_{p}+2
\end{array}
\right]
\end{equation}

\end{enumerate}

The particular form of the matrix $K$ and the weak rank of the matrix $M$ allows
an explicit calculation of the eigenvalues in all the cases $p=6,7,8$ and for all
the Bianchi Universes. There exists several hundreds of non vanishing minor
determinants in all the cases considered. They correspond to 92 distinct sets
of Kovalewski exponents with length $2p=12,14$ or $16$. All the details are
presented in appendix B. In all cases, calculations are explicit, and in order to organize 
it we use formal calculus computational tools.

\subsubsection{Exponents analysis and conclusions}

The full set of Kovalewski exponents is presented in appendix B. The analysis of
the sets of exponents results in four classes for Bianchi models :  Class I
contains uniquely $B_{\mbox{\textsc{I}}}$ models, Class II contains
$B_{\mbox{\textsc{II}}}$ and $B_{\mbox{\textsc{IV}}}$, Class III is composed
by  $B_{\mbox{\textsc{III}}}$, $B_{\mbox{\textsc{VI}}_{o,a}}$ and
$B_{\mbox{\textsc{VII}}_{o,a}}$, and finally Class IV contains $B_{\mbox{\textsc{VIII}}}$ and the famous $B_{\mbox{\textsc{IX}}}$. Each class is characterized by the same set of Kovalewski exponents, therefore corresponding models have the same characteristic dynamics. Let us review for each class the algebraic properties of Kovalewski exponents :

\begin{itemize}
\item Class I $\;-\;$
For all the cases we studied, all the Kovalewski exponents belonging to this class, are rational provided that the barotropic index $\Gamma$ and/or the scalar-tensor parameter $\Delta$ are rational. This restriction corresponds to physical cases\footnote{ As a matter of fact, physical power laws or barotropic index must be rational in order to be full of physical meaning}.

\item Class II $\;-\;$
Kovalewski exponents of models belonging to this class fall into four subcases :
\begin{itemize}
\item Empty Universe in General Relativity (EUGR): exponents belonging to this class are integers.
\item Barotropic Matter filled Universe in General Relativity (BMUGR): due to 
$A_{\pm}^{*}$, all exponents are rational iff 
$\Gamma \in \mathbb{Q}\cap[0,\Gamma_0]$ with $\Gamma_0=(11-\sqrt{73})/3\approx0.81$.  
\item Empty Universe in Scalar-Tensor Theory (EUSTT): due to 
$A_{\pm}$, all exponents are rational iff 
$\Delta \in \mathbb{Q}\cap[\Delta_0,1]$ with $\Delta_0=(-11+\sqrt{73})/6\approx-0.40$.  
\item Barotropic Matter filled Universe in Scalar-Tensor Theory (BMUSTT): Due to 
$A_{\pm}$ and $D_{\pm}$, all exponents are rational iff 
$\Delta \in \mathbb{Q}\cap[\Delta_1,1]$ with $\Delta_1=(-5+\sqrt{73})/6\approx0.59$.  
\end{itemize}

\item Class III $\;-\;$
As previously, four classes of sets of Kovalewski exponents can be identified
\begin{itemize}
\item  EUGR : exponents belonging to this class are integers.
\item  BMUGR : due to 
$A_{\pm}^{*}$, $B_{\pm}^{*}$ and $C_{\pm}^{*}$, all exponents are rational iff 
$\Gamma \in \mathbb{Q}\cap[0,\Gamma_0]$. This case is then equivalent to the corresponding case of class II models. Let us remark the special value $\Gamma=2/3$ for which $(A_{+}^{*},\;A_{-}^{*},\;B_{+}^{*},\;B_{-}^{*},\;C_{+}^{*},\;C_{-}^{*})=(1,\;0,\;1,\;0,\;1,\;0)$. 
\item  EUSTT : due to 
$A_{\pm}$,$B_{\pm}$ and $a_{\pm}$ all exponents are rational iff 
$\Delta \in \mathbb{Q}\cap[\Delta_0,1]$. This case is then equivalent to the corresponding case of class II models. Let us remark the special value $\Delta=-1/3$ for which $(A_{+},\;A_{-},\;B_{+},\;B_{-},\;a_{+},\;a_{-})=(1,\;0,\;1,\;0,\;1,\;0)$.  
\item  BMUSTT : due to 
$A_{\pm}$,$B_{\pm}$, $C_{\pm}$, $D_{\pm}$, $a_{\pm}$ and $b_{\pm}$, all exponents are rational iff 
$\Delta \in \mathbb{Q}\cap[\Delta_2,1]$ with $\Delta_2=16/25=0.64$.  
\end{itemize}

\item Class IV $\;-\;$  all models of this class contain at least two conjugated complex Kovalewski exponents.
\end{itemize}
>From such an analysis, taking into account Yoshida's theorems (see \cite{Yoshida1,Yoshida2}) whose context is detailed in appendix A, we can conclude that:
\begin{itemize}
\item Empty Universes whose metrics correspond to classes I, II or III defined below are generically associated to algebraically integrable dynamics.
\item Class IV Universes are always associated to non algebraically integrable
  dynamics. This result holds for General Relativity and/or scalar-tensor
  theory we are interested in, the Universe being empty and/or filled of barotropic matter. 
\item BMUGR of class II and III are associated to non integrable dynamics if the barotropic index ranges in the interval $[\Gamma_0,2]$ with $\Gamma_0=(11-\sqrt{73})/3\approx0.81$.
\item EUSTT of class II and III are associated to non integrable dynamics if
  the power law of the scalar-tensor modelization (\ref{vdelta}) ranges in
  the interval $[-2,\Delta_0]$ with
  $\Delta_0=(-11+\sqrt{73})/6\approx-0.40$.
\item BMUSTT of class II are associated to non integrable dynamics if the power law of the scalar tensor modelization (see \ref{vdelta}) ranges in the interval $[-2,\Delta_1]$ with $\Delta_1=(-5+\sqrt{73})/6\approx-0.40$.
\item BMUSTT of class III are associated to non integrable dynamics if the power
  law of the scalar tensor modelization (\ref{vdelta}) ranges in the
  interval $[-2,\Delta_2]$ with $\Delta_2=16/25=0.64$.
\end{itemize}

The integrability of homogeneous Universes depends on multiple factors. 

It is well known, since the pioneering works by  Elskens and Henneaux \cite{EetH}, that the number of
dimensions $D$ of the spatial sections of the Universe is one of this
factors. It influences the form of the potential $\xi \left( q\right) \ \ $%
in relation $\left( \ref{xiqq}\right) $ or/and increases the number of
components of the vectors  $\left\{ \mathbf{a}_{i}\right\} $ in $\left( \ref%
{liste}\right) $. The net result in the simplified context of generalized
Kasner metric is that chaos disappears $-$ and then system becomes always
integrable, in a classical context $-$ when $D\geq 11.$ 

Modification of the gravitation theory by introducing a scalar-tensor
component seems here to have a different contribution. As a matter of fact,
in the case we consider, the potential  $\xi \left( q\right) $ is not
modified, but a volume dependent Toda potential is added in the Hamiltonian. In a general way, this produces a slowing
down for the cushions' velocity  in the billiard representation (see
\cite{Jantzen} and references therein). One can then reasonably suppose that
this feature does not change drastically the dynamical properties of
homogeneous Universes inherited from General Relativity. The integrability analysis based on
Kovalewski exponents we have done confirms such a conjecture. 

\appendix
\section*{Appendix A : Integrability of autosimilar differential systems}
\setcounter{section}{1}
Let $\mathbf{f}$ be a continuous function from $\mathbb{R}^{n}$ to $\mathbb{R}%
^{n}$ and $\mathbf{x}=\left[  x_{1}(t),...,x_{n}(t)\right]  ^{T}\in
\mathbb{R}^{n}$ such that
\begin{equation}
\frac{d\mathbf{x}}{dt}=\mathbf{f}\left(  \mathbf{x}\right)  \label{sysdynaut}%
\end{equation}
If there exists a positive real $\lambda$ and a rational vector $\mathbf{g}%
:=\left[  g_{1},...,g_{n}\right]  ^{T}$ such that the transformation
\begin{equation}%
\begin{array}
[c]{c}%
t\mapsto t/\lambda\\
\forall i=1,...,n~\ \ x_{i}\mapsto\lambda^{g_{i}}x_{i}%
\end{array}
\label{critautosim}%
\end{equation}
leaves the system $\left(  \ref{sysdynaut}\right)  $ \ invariant, this system is
called autosimilar with index $\lambda$ and weight $\mathbf{g}$.

When it exists $\mathbf{g}$, it is unique if
\begin{equation}
A\left(  \mathbf{x}\right)  \in M_{n}\left(  \mathbb{R}\right)  \mbox{ ,
\ \ }A_{ij}=x_{j}\frac{\partial f_{i}\left(  \mathbf{x}\right)  }{\partial
x_{j}}-\delta_{ij}f_{i}\left(  \mathbf{x}\right)  \label{condunicite}%
\end{equation}
is invertible for almost all $\mathbf{x\in}\mathbb{R}^{n}$.

Autosimilar systems always admit at least one autosimilar particular solution
\begin{equation}
\mathbf{\tilde{x}}_{as}=\left[  c_{1}\left(  t-t_{o}\right)  ^{-g_{1}%
},...,c_{n}\left(  t-t_{o}\right)  ^{-g_{n}}\right]  ^{T}
\end{equation}
where
$\mathbf{c}:=\left[  c_{1},...,c_{n}\right]^{T}$is solution of the algebraic
equation:
\begin{equation}
\left\{
\begin{array}
[c]{c}%
f_{1}\left(  \mathbf{c}\right)  =-g_{1}c_{1}\\
\vdots\\
f_{n}\left(  \mathbf{c}\right)  =-g_{n}c_{n}%
\end{array}
\right.  \label{SNL}%
\end{equation}
Linearization of the system $\left(  \ref{sysdynaut}\right)  $ around the
solution$\ \mathbf{\tilde{x}}_{as}$ gives
\begin{equation}
\frac{d\mathbf{z}}{dt}=D\mathbf{f}\left(  \mathbf{x}\right)  \left(
\mathbf{\tilde{x}}_{as}\right)  \ \mathbf{z} \label{linovoiz}%
\end{equation}
where $D\mathbf{f}\left(  \mathbf{x}\right)  \left(  \mathbf{\tilde{x}}%
_{as}\right)  $ is the Jacobian matrix of $\mathbf{f}\left(  \mathbf{x}%
\right)  $ evaluated at $\mathbf{x=}$ $\mathbf{\tilde{x}}_{as}$. A theorem by
Fuchs (see \cite{conte} \S 5.6 and 5.7) then shows that the general solution of $\left(  \ref{linovoiz}\right)  $ is
autosimilar and reads:
\begin{equation}
\mathbf{z}=\left[  k_{1}\left(  t-t_{o}\right)  ^{\rho_{1}-g_{1}}%
,...,k_{n}\left(  t-t_{o}\right)  ^{\rho_{n}-g_{n}}\right]  ^{T}%
\end{equation}
The quantities $\left(  \rho_{1},\rho_{2},...,\rho_{n}\right)  $ are called
Kovalewski exponents. More practically, one can compute them because there are
also eigenvalues of the matrix:
\begin{equation}
K:=D\mathbf{f}\left(  \mathbf{x}\right)  \left(  \mathbf{c}\right)
+\mathrm{diag}\left(  \mathbf{g}\right)  . \label{kowmat}%
\end{equation}

As shown by Comte (see \cite{conte} \S 5.6 and 5.7), they are of
great importance for the study of integrability of the original non linear
system $\left(  \ref{sysdynaut}\right)  $. As a matter of fact each component
$x_{i=1,..,n}$ of a solution $x$ of this system can be written as:
\begin{equation}
\fl
x_{i}\left(  t\right)  =\sum_{k=0}^{+\infty}\varepsilon^{k}x_{i}^{\left(
k\right)  }\left(  t\right)  \ \ \ \ \ \ \mbox{with\ \ \ \ \ }\left\{
\begin{array}
[c]{l}%
x_{i}^{\left(  0\right)  }=c_{i}\left(  t-t_{o}\right)  ^{g_{i}}\\
x_{i}^{\left(  1\right)  }=k_{i}\left(  t-t_{o}\right)  ^{\rho_{i}-g_{i}}\\
\exists\ 1\leq p,q\leq n\mbox{, \ \ }x_{i}^{\left(  2\right)  }=c_{p}%
k_{q}\left(  t-t_{o}\right)  ^{g_{p}+\rho_{q}-g_{q}}\\
\mbox{etc }\cdots
\end{array}
\right.
\end{equation}
hence
\begin{equation}
x_{i}\left(  t\right)  \propto\left(  t-t_{o}\right)  ^{g_{i}}S\left[  \left(
t-t_{o}\right)  ^{\rho_{1}},...,\left(  t-t_{o}\right)  ^{\rho_{n}}\right]
\label{serie}%
\end{equation}
where $S\left[  \;\,\right]  $ is a multiple series. This allows to understand a
theorem by Yoshida (see \cite{Yoshida1,Yoshida2}) : a necessary
condition for a differential system to be integrable is that all its
Kovalewski exponents are rational. If there exists at least one exponent
irrational or complex the corresponding differential system is not algebraically integrable.

\appendix
\section*{Appendix B : Details of Kovalewski exponents}
\setcounter{section}{1}
When only non exotic barotropic matter ( $0\leq\Gamma<2)$ is considered, we
have noted $\gamma=(3\,\Gamma-2)/(\Gamma-2)$. For any $K=3,5$,$7$ and 15 we
note $2K^{\pm i}=\,1\pm i\sqrt{K}$ and
\[%
\begin{array}
[c]{cc}%
2A_{\pm}^{\ast}=1\pm\sqrt{\frac{16-22\,\Gamma+3\,\Gamma^{2}}{4-2\Gamma}} &
2B_{\pm}^{\ast}={1}\pm\,\sqrt{5-6\,\Gamma}\\
2C_{\pm}^{\ast}=1\pm\sqrt{\frac{25\Gamma-18}{\Gamma-2}} & 2D_{\pm}^{\ast}%
=1\pm\sqrt{\frac{18-41\,\Gamma+24\,\Gamma^{2}}{2-\Gamma}}%
\end{array}
\]%
\[%
\begin{array}
[c]{cc}%
E_{\gamma}:=\left\{  -1,2,\,\,1\left(  \times6\right)  ,\frac{2{\gamma}}%
{3}\left(  \times6\right)  \right\}  & E_{o}:=\left\{  {-1,4(\times
3),2(\times3),1(\times5)}\right\}
\end{array}
\]

Moreover%
\[
E_{\delta}=[\left\{  -1,1\;\mbox{{\small (}}{\small \times}\mbox{{\small 6)\ }%
},2\,\,,\,2\delta_{1}/3\mbox{{\small (}}{\small \times}\mbox{{\small 6)}%
}\right\}
\]

\[
\delta_{1}=\left(  3\Delta+1\right)\left(  \Delta+1\right)  ^{-1}\ \ \ \ \ \mbox{and\ \ \ \ \ \ \ }%
\delta_{2}=\left(  3\,\Delta-2\right)  \Delta^{-1}%
\]

$\;\;$%

\[
2A_{\pm}{=}1\pm\,\sqrt{\frac{3\left(  \Delta-\Delta_{A}^{+}\right)  \left(
\Delta-\Delta_{A}^{-}\right)  }{\left\vert \Delta+1\right\vert }%
}\ \ \ \ \ \ \ \ \ 6\Delta_{A}^{\pm}=-11\pm\sqrt{73}\ \
\]

\[
2D_{\pm}=1\pm\,\sqrt{\frac{3\left(  \Delta-\Delta_{D}^{+}\right)  \left(
\Delta-\Delta_{D}^{-}\right)  }{\left\vert \Delta\right\vert }}%
\ \ \ \ \ \ \ \ \ \ \ \ 6\Delta_{D}^{\pm}=-5\pm\sqrt{73}%
\]

${\;}$%

\[
2B_{\pm}=1\pm\,\sqrt{\frac{25\Delta+9}{\left\vert \Delta+1\right\vert }}%
\]

${\ \ \ }$%

\[
2C_{+}=1\pm\,\sqrt{\frac{25\Delta-16}{\left\vert \Delta\right\vert }}%
\]

${\;\;}$%

\[
2E_{\pm}=1\pm\,\sqrt{\frac{48\Delta^{2}+41\Delta+9}{\left\vert \Delta
+1\right\vert }}\ \ \ \ \ \ \ \ \ \ \ \ \ 48\Delta^{2}+41\Delta+9>0
\]

\[
2F_{\pm}=1\pm\,\sqrt{\frac{48\Delta^{2}-55\Delta+16}{\left\vert \Delta
\right\vert }}\ \ \ \ \ \ \ \ \ \ 48\Delta^{2}-55\Delta+16>0
\]

\bigskip%

\[
2a_{\pm}=1\pm\,\sqrt{5+12\,\Delta}\ \ \ \ \ \ \ \ \ \ 2b_{\pm}=1\pm
\,\sqrt{-7+12\,\Delta}%
\]

\bigskip Using the notations defined above we have computed all the Kowalewski exponents
for all Bianchi Universes, in the case of empty or barotropic filled Universes
and for General Relativity(GR) and Scalar-Tensor (ST) theory of gravitation:

\scriptsize{\begin{tabular}
[c]{|ccc|}\hline\hline
$B_{\mbox{\textsc{i}}}$ &  & \\\hline\hline
RG & \multicolumn{1}{|c}{{\small Empty (N=6)}} &
\multicolumn{1}{|c|}{{\small Barotropic Matter (N=7)}}\\\hline\hline
{\small T1} & \multicolumn{1}{|c}{$\varnothing$} &
\multicolumn{1}{|c|}{$E_{\gamma}$}\\\hline
{\small T2} & \multicolumn{1}{|c}{$\varnothing$} &
\multicolumn{1}{|c|}{$\varnothing$}\\\hline
{\small T3} & \multicolumn{1}{|c}{$\varnothing$} &
\multicolumn{1}{|c|}{$\varnothing$}\\\hline\hline
ST & \multicolumn{1}{|c}{{\small Empty (N=7)}} &
\multicolumn{1}{|c|}{{\small Barotropic Matter (N=8)}}\\\hline\hline
{\small T1} & \multicolumn{1}{|c}{$E_{\delta}$} & \multicolumn{1}{|c|}{$\begin{array}
[c]{c}-1,\,,\,1\mbox{{\small (}}{\small \times}\mbox{{\small 7), }}2\;,\;2\delta
_{2}/3\mbox{{\small (}}{\small \times}\mbox{{\small 6)}},\,-2/\Delta\\
-1,\,1\mbox{{\small (}}{\small \times}\mbox{{\small 7)}},2\,,{2\left(  \Delta+1\right)  ^{-1}},\,2\delta_{1}/3\mbox{{\small (}}{\small \times}\mbox{{\small 6)}}\end{array}
$}\\\hline
{\small T2} & \multicolumn{1}{|c}{$\varnothing$} &
\multicolumn{1}{|c|}{$\varnothing$}\\\hline
{\small T3} & \multicolumn{1}{|c}{$\varnothing$} &
\multicolumn{1}{|c|}{$\varnothing$}\\\hline\hline
\end{tabular}
}

\begin{tabular}
[c]{|ccc|}\hline\hline
$B_{\mbox{\textsc{ii}}}${\small \ and }$B_{\mbox{\textsc{iv}}}$ &  &
\\\hline\hline
GR & \multicolumn{1}{|c}{{\small Empty (N=6)}} &
\multicolumn{1}{|c|}{{\small Barotropic Matter (N=7)}}\\\hline\hline
{\small T1} & \multicolumn{1}{|c}{$E_{o}$} & \multicolumn{1}{|c|}{%
\begin{tabular}
[c]{l}%
$E_{\gamma}$\\
$-1,1(\times6),\,\,2(\times3),\,4(\times3),3-\Gamma/2\,\,$%
\end{tabular}
}\\\hline
{\small T2} & \multicolumn{1}{|c}{$\varnothing$} &
\multicolumn{1}{|c|}{$-1,1\left(  \times5\right)  ,2,\gamma/2\left(
\times2\right)  ,\gamma\left(  \times3\right)  ,A_{\pm}^{\ast}$}\\\hline
{\small T3} & \multicolumn{1}{|c}{$\varnothing$} &
\multicolumn{1}{|c|}{$\varnothing$}\\\hline\hline
ST & \multicolumn{1}{|c}{{\small Empty (N=7)}} &
\multicolumn{1}{|c|}{{\small Barotropic Matter (N=8)}}\\\hline\hline
{\small T1} & \multicolumn{1}{|c}{$%
\begin{array}
[c]{l}%
-1,\,1{\small (\times6)},\,\,2{\small (\times3)},\,4{\small (\times3)}%
,\Delta+3\\
E_{\delta}%
\end{array}
$} & \multicolumn{1}{|c|}{$%
\begin{array}
[c]{l}%
-1,\,1{\small (\times7)},2,\,\,2\,{\left(  \Delta+1\right)  ^{-1}},2\delta
_{1}/3\,{\small (\times6)}\\
-1,\,1{\small (\times7)},\,2,2\delta_{2}/3{\small (\times6)},\,-2\,/\Delta\\
-1,\,1{\small (\times7)},\,2{\small (\times3)},\,4{\small (\times3)}%
,\,\Delta+3,\,\Delta+2
\end{array}
$}\\\hline
{\small T2} & \multicolumn{1}{|c}{$-1,\,1{\small (\times5)},\,2,\,\,{\delta
_{1}/2{\small (\times2)}},\,{\delta_{1}{\small (\times3)}},\,A_{\pm}$} &
\multicolumn{1}{|c|}{$%
\begin{array}
[c]{l}%
-1,\,1{\small (\times6)},\,2,\,{\delta_{2}/2{\small (\times2)}},\,{\delta
_{2}{\small (\times3)}}\,,-2/\Delta,\,D_{\pm}\\
-1,\,1{\small (\times6)},\,2,\,\,{\delta_{1}/2{\small (\times2)}}%
,\,{\delta_{1}{\small (\times3)}},\,2\,{\left(  \Delta+1\right)  ^{-1}}%
,\,{A}_{\pm}%
\end{array}
$}\\\hline
{\small T3} & \multicolumn{1}{|c}{$\varnothing$} &
\multicolumn{1}{|c|}{$\varnothing$}\\\hline\hline
\end{tabular}

\begin{tabular}[c]{|ccc|}\hline\hline
$B_{III}$ ,
$B_{VI_{o,a}}$, 
$B_{VII_{o,a}}$

 &  & \\ \hline\hline
GR & \multicolumn{1}{|c}{{\small Empty (N=6)}} &
\multicolumn{1}{|c|}{{\small Barotropic Matter(N=7)}}\\\hline\hline
{\small T1} & \multicolumn{1}{|c}{$E_{o}$} & \multicolumn{1}{|c|}{%
\begin{tabular}
[c]{l}%
$E_{\gamma}$\\
$-1\,,\,1(\times6),\,\,2(\times3),\,4(\times3),3-\Gamma/2$%
\end{tabular}
}\\\hline
{\small T2} & \multicolumn{1}{|c}{$\varnothing$} & \multicolumn{1}{|c|}{%
\begin{tabular}
[c]{l}%
$-1,1\left(  \times5\right)  ,2,\gamma/2\left(  \times2\right)  ,\gamma\left(
\times3\right)  ,A_{\pm}^{\ast}$\\
$-1,\,0(\times2),\,1(\times5),\,2,\,\,2\,\gamma,\,\gamma(\times2),\,B_{\pm
}^{\ast}$%
\end{tabular}
}\\\hline
{\small T3} & \multicolumn{1}{|c}{$\varnothing$} & \multicolumn{1}{|c|}{%
\begin{tabular}
[c]{l}%
$-1,\,0,\,1(\times4),\,2,\,2\,\gamma,\,\gamma(\times2),\,B_{\pm}^{\ast
},\,\,C_{\pm}^{\ast}$\\
$-1,\,0(\times2),\,1(\times5),\,2,\,\,2\,\gamma,\,\gamma(\times2),\,B_{\pm
}^{\ast}$%
\end{tabular}
}\\\hline\hline
ST & \multicolumn{1}{|c}{{\small Empty (N=7)}} &
\multicolumn{1}{|c|}{{\small Barotropic Matter (N=8)}}\\\hline\hline
{\small T1} & \multicolumn{1}{|c}{$%
\begin{array}
[c]{l}%
{-1,\,1{\small (\times6)},\,\,2{\small (\times3)},\,4{\small (\times3)}%
,\Delta+3}\\
E_{\delta}%
\end{array}
$} & \multicolumn{1}{|c|}{$%
\begin{array}
[c]{l}%
-1,\,1{\small (\times7)},2,\,2\delta_{1}/3{\small (\times6)},\,\,2\,{\left(
\Delta+1\right)  ^{-1}}\\
-1,\,1{\small (\times7)},\,2\,,2\delta_{2}/3{\small (\times6)},\,-2\Delta
^{-1}\\
-1,\,1{\small (\times7)},\,\,2{\small (\times3)},\,4{\small (\times3)}%
,\Delta+3,\,\Delta+2
\end{array}
$}\\\hline
{\small T2} & \multicolumn{1}{|c}{$%
\begin{array}
[c]{l}%
-1,\,1{\small (\times5)},\,2\;,\,{\delta_{1}/2\;{\small (\times2)}%
},\,\,{\delta_{1}{\small (\times3)}},\,A_{\pm}\\
-1,\,0{\small (\times2)}\,,1{\small (\times5)},\,2,2\,{\delta_{1}}%
,\,{\delta_{1}{\small (\times2)}},\,a_{\pm}%
\end{array}
$} & \multicolumn{1}{|c|}{$%
\begin{array}
[c]{l}%
-1,\,0{\small (\times2)},\,1{\small (\times6)},\,2,\,2\,{\delta_{1}}%
,\,{\delta_{1}{\small (\times2)}},\,2\,{\left(  \Delta+1\right)  ^{-1}%
}\,,a_{\pm}\\
-1,\,1{\small (\times6)},\,2,\,\,{\delta_{1}/2\;{\small (\times2)}}%
,{\delta_{1}{\small (\times3)}},\,2\,{\left(  \Delta+1\right)  ^{-1}}%
,\,{A}_{\pm}\\
-1,\,0{\small (\times2)},\,1{\small (\times6)},2,\,2\,{\delta_{2}}%
,\,{\delta_{2}{\small (\times2)}},-2\,\Delta^{-1},\,b_{\pm}\\
-1,\,1{\small (\times6)}\,,\,\,2,\,\,{\delta_{2}/2{\small (\times2)}%
},\,{\delta_{2}{\small (\times3)}},-2\Delta^{-1},\,D_{\pm}%
\end{array}
$}\\\hline
{\small T3} & \multicolumn{1}{|c}{$%
\begin{array}
[c]{l}%
-1,\,0,\,1{\small (\times4)},\,2,\,2\,{\delta_{1}},\,{\delta}_{1}%
{\small (\times2)},\,{a}_{\pm},\,B_{\pm}\\
-1,\,0{\small (\times2)},\,1{\small (\times5)}\,,2,\,2\,{\delta_{1}}%
,\,{\delta_{1}{\small (\times2)}},\,{a}_{\pm}%
\end{array}
$} & \multicolumn{1}{|c|}{$%
\begin{array}
[c]{l}%
-1,\,0,\,1{\small (\times5)}\,,2,\,2\,{\delta_{1}},\,{\delta_{1}%
{\small (\times2)}},\,2\,{\left(  \Delta+1\right)  ^{-1}},\,a_{\pm},\,B_{\pm
}\\
-1,\,0{\small (\times2)},\,1{\small (\times6)},\,2,\,2\,{\delta_{1}}%
,\,{\delta_{1}{\small (\times2)}},\,2\,{\left(  \Delta+1\right)  ^{-1}%
},\,a_{\pm}\\
-1,\,0{\small (\times2)},\,1{\small (\times6)},\,\,2,2\,{\delta_{2}}%
,{\delta_{2}{\small (\times2)}},\,-2\Delta\,^{-1},\,b_{\pm}\\
-1,0,\,1{\small (\times5)},\,\,2,{\delta_{2}{\small (\times2)}},\,2\,{\delta
_{2}},-2\Delta\,^{-1},b_{\pm},\,C_{\pm}%
\end{array}
$}\\\hline\hline
\end{tabular}

\bigskip%

\begin{tabular}
[c]{|ccc|}\hline\hline
$B_{\mbox{\textsc{viii}}}${\small \ and }$B_{\mbox{\textsc{ix}}}$ &  &
\\\hline\hline
GR & {\small Empty} {\small (N=6)} & {\small Barotropic Matter (N=7)}%
\\\hline\hline
{\small T1} & $E_{o}$ &
\begin{tabular}
[c]{l}%
$E_{\gamma}$\\
$-1,\,1(\times6),\,\,2(\times3),\,4(\times3),3-\Gamma/2\,$%
\end{tabular}
\\\hline
{\small T2} &
\begin{tabular}
[c]{l}%
$-1,-2(\times3),7^{\pm i},\,2(\times2),\,\,1(\times4)$\\
$-1,\,2,\,15^{\pm i},\,1(\times4),\,0(\times4)$%
\end{tabular}
&
\begin{tabular}
[c]{l}%
$-1,1\left(  \times5\right)  ,2,\gamma/2\left(  \times2\right)  ,\gamma\left(
\times3\right)  ,A_{\pm}^{\ast}$\\
$-1,\,0(\times2),\,1(\times5),\,2,\,\,2\,\gamma,\,\gamma(\times2),\,B_{\pm
}^{\ast}$\\
$-2(\times3),-1,\,\,1(\times5),\,2(\times2),7^{\pm i},2\Gamma-2$\\
$-1,\,0(\times4),\,1(\times5),\,2,\,\,15^{\pm i},-1+3\,\Gamma/2$%
\end{tabular}
\\\hline
{\small T3} & $%
\begin{tabular}
[c]{l}%
$-1,2,\,\,15^{\pm i}(\times2),\,1(\times3),\,0(\times3)$\\
$-1,2,\,\,3^{\pm i}(\times2),\,\ \ 1(\times3),\,0(\times3)$\\
$-1,\,2,\,15^{\pm i},\,1(\times4),\,0(\times4)$\\
$-1(\times2),2(\times2),7^{\pm i},\,1(\times3),\,0(\times3)$%
\end{tabular}
\ \ $ &
\begin{tabular}
[c]{l}%
$-1,\,0(\times3),\,1(\times4),\,2,3^{\pm i}(\times2),-1+\,3\Gamma/2$\\
$-1,\,1(\times4),\,2,\,-2\,\gamma,2\,\gamma(\times3),C_{\pm}^{\ast},D_{\pm
}^{\ast}$\\
$-1,\,0(\times3),\,1(\times4),\,2,\,\,15^{\pm i}(\times2),-1+3\,\Gamma/2$\\
$-1,\,0(\times4),\,1(\times5),\,2,\ 15^{\pm i},-1+3\,\Gamma/2$\\
$-1,\,0,\,1(\times4),\,2,\,2\,\gamma,\,\gamma(\times2),\,B_{\pm}^{\ast
},\,\,C_{\pm}^{\ast}$\\
$-1(\times2),\,\,0(\times3),\,1(\times4),\,2(\times2),\,7^{\pm i}%
,-1+3\,\Gamma/2$\\
$-1,\,0(\times2),\,1(\times5),\,2,\,\,2\,\gamma,\,\gamma(\times2),\,B_{\pm
}^{\ast}$%
\end{tabular}
\\\hline\hline
ST & {\small Empty (N=7)} & {\small Barotropic Matter (N=8)}\\\hline\hline
{\small T1} & $%
\begin{array}
[c]{l}%
E_{\delta}\\
-1,\,1{\small (\times6)},\,\,2{\small (\times3)},\,4{\small (\times3)}%
,\Delta+3
\end{array}
$ & $%
\begin{array}
[c]{l}%
-1,\,1{\small (\times7)},\,2,2\delta_{2}/3{\small (\times6)},\,-2\Delta
^{-1}\,\\
-1,\,1{\small (\times7)},\,2{\small (\times3)},\,\,4{\small (\times3)}%
,\Delta+3,\,\Delta+2\\
-1,\,1{\small (\times7)},2,2\delta_{1}/3\,{\small (\times6)},\,\,2\,{\left(
\Delta+1\right)  ^{-1}}%
\end{array}
$\\\hline
{\small T2} & $%
\begin{array}
[c]{l}%
-1,\,0{\small (\times2)},\,1{\small (\times5)},\,2,\,{\delta_{1}%
}{\small (\times2)},\,2\,{\delta_{1}},\,a_{\pm}\\
-2{\small (\times3)},-1,\,\,1{\small (\times5)},\,2{\small (\times2)},\,7^{\pm
i},-4\,\Delta-2\\
-1,\,0{\small (\times4)},\,1{\small (\times5)},\,\,2,\,{15}^{\pm i}%
,-3\,\Delta-1\\
-1,\,1{\small (\times5)},\,2,{\delta_{1}/2}\,{\small (\times2)},\,{\delta_{1}%
}{\small (\times3)},{A}_{\pm}%
\end{array}
$ & $%
\begin{array}
[c]{l}%
-1,\,1{\small (\times6)},\,\,2,\,{D}_{\pm},\,{\delta}_{2}/2{\small (\times
2)},\,{\delta_{2}}{\small (\times3)},-2\,\Delta^{-1}\\
-1,\,0{\small (\times2)},\,1{\small (\times6)},\,\,2,2\,{\delta_{2}}%
,\,{\delta_{2}}{\small (\times2)},\,-2\,\Delta^{-1},\,b_{\pm}\\
-1,\,0{\small (\times2)},\,1{\small (\times6)},\,2,\,2\,{\delta_{1}}%
,\,{\delta_{1}}{\small (\times2)},2\,{\left(  \Delta+1\right)  ^{-1}}%
,\,a_{\pm}\\
-1,\,1(\times6),\,2,\,\,\delta_{1}/2(\times2),\delta_{1}(\times3),\,2\,\left(
\Delta+1\right)  ^{-1},A_{\pm}\\
-2(\times3),-1,\,\,1(\times6),\,2(\times2),\,\,7^{\pm i},-4\,\Delta
-2,\,-4\,\Delta+2\\
-1,\,0{\small (\times4)},\,1{\small (\times6)},\,\,2,\,{15}^{\pm i}%
,-3\,\Delta-1,\,-3\,\Delta+2
\end{array}
$\\\hline
{\small T3} & $%
\begin{array}
[c]{l}%
-1,\,0\,{\small (\times2)},\,1\,{\small (\times5)},\,2,\,2\,{\delta_{1}%
},\,{\delta_{1}}\,{\small (\times2)},\,{a}_{\pm}\\
-1,\,0\,{\small (\times4)},\,1\,{\small (\times5)},\,\,2,\,{15}^{\pm
i},-3\,\Delta-1\\
-1,\,0\,{\small (\times3)},\,1\,{\small (\times4)},\,2,{15}^{\pm
i}\,{\small (\times2)},\,-3\,\Delta-1\\
-1,\,0{\small (\times3)},\,1{\small (\times4)},\,2,\,\,{3}^{\pm i}%
{\small (\times2)},-3\,\Delta-1\\
-1,\,0,\,1{\small (\times4)},\,2,\,2\,{\delta_{1}},{\delta_{1}}{\small (\times
2)},{a}_{\pm},\,{B}_{\pm}\\
-1{\small (\times2)},\,\,0{\small (\times3)},\,1{\small (\times4)}%
\,{\small ,}2{\small (\times2)},\,{7}^{\pm i},-3\,\Delta-1\\
-1,\,1{\small (\times4)},\,2,-2\,{\delta_{1}},\,2\,{\delta_{1}}{\small (\times
3)},\,{B}_{\pm},E_{\pm}%
\end{array}
$ & $%
\begin{array}
[c]{l}%
-1,\,1{\small (\times5)},\,2,\,-2\,{\delta_{1}},\,2\,{\delta_{1}%
}{\small (\times3)},2\,{\left(  \Delta+1\right)  ^{-1}},\,{B}_{\pm},\,{E}%
_{\pm}\\
-1,\,1{\small (\times5)},\,2,-2\,{\delta_{2}},\,2\,{\delta_{2}}{\small (\times
3)},-2\,\Delta^{-1},\,{C}_{\pm},\,{F}_{\pm}\\
-1,\,0{\small (\times3)},\,1{\small (\times5)},\,2,\,\,3^{\pm i}%
{\small (\times2)},-3\,\Delta-1,-3\,\Delta+2\\
-1,\,0{\small (\times3)},\,1{\small (\times5)},\,2,\,\,{15}^{\pm
i}{\small (\times2)},-3\,\Delta-1,\,-3\,\Delta+2\\
-1{\small (\times2)},\,\,0{\small (\times3)},\,1{\small (\times5)}%
,\,2{\small (\times2)},\,{7}^{\pm i},\,-3\,\Delta-1,\,-3\,\Delta+2\\
-1,\,0,\,1{\small (\times5)},2,\,\ 2\,{\delta_{1}},\ {\delta_{1}%
}{\small (\times2)},\,2\,{\left(  \Delta+1\right)  ^{-1}},\,{a}_{\pm}%
,\,{B}_{\pm}\\
-1,0,\,\ 1{\small (\times5)},\,\,2,\,2\,{\delta_{2}},\,{\delta_{2}%
}{\small (\times2)},-2\Delta^{-1},\,{b}_{\pm},\,{C}_{\pm}\\
-1,\,0{\small (\times2)},\,1{\small (\times6)},\,\,2,2\,{\delta_{2}}%
,\,{\delta_{2}}{\small (\times2)},-2\Delta^{-1},\,{b}_{\pm}\\
-1,\,0{\small (\times2)},\,1{\small (\times6)},\,2,\,2\,{\delta_{1}}%
,\,{\delta_{1}}{\small (\times2)},\,2\,{\left(  \Delta+1\right)  ^{-1}}%
,\,{a}_{\pm}\\
-1,\,0{\small (\times4)},\,1{\small (\times6)},\,\,2,\,{15}^{\pm i}%
,-3\,\Delta-1,\,-3\,\Delta+2
\end{array}
$\\\hline
\end{tabular}


\end{document}